# Camouflaged with Size: A Case Study of Espionage using Acquirable Single-Board Computers

Kiavash Satvat[1], Mahshid Hosseini [1] and Maliheh Shirvanian[2]

[1]Department of Computer Science, University of Illinois at Chicago, Chicago, USA
ksatva2, mhosse4@uic.edu
[2]Department of Computer Science, University of Alabama at Birmingham, Alabama, USA
maliheh@uab.edu

## ABSTRACT

*Single-Board Computers (SBC) refer to pocket-sized computers built on a single circuit board. A number of studies have explored the use of these highly popular devices in a variety of domains, including military, agriculture, healthcare, and more. However, no attempt was made to signify possible security risks that misuse of these devices may bring to organizations. In this study, we perform a series of experiments to validate the possibility of using SBCs as an espionage gadget. We show how an attacker can turn a Raspberry Pi device to an attacking gadget and benefit from short-term physical access to attach the gadget to the network in order to access unauthorized data or perform other malicious activities. We then provide experimental results of placing such tools in two real-world networks. Given the small size of SBCs, traditional physical security measures deployed in organizations may not be sufficient to detect and restrict the entrance of SBCs to their premises. Therefore, we reiterate possible directions for network administrators to deploy defensive mechanisms for detecting and preventing such attacks.*

## KEYWORDS

*Espionage, Single-Board Computer (SBC), Physical Security, Network Security, Raspberry Pi.*

## 1. INTRODUCTION

Single-Board Computer (SBC) is a pocket-sized computer that is built on a single circuit board. Using microprocessors and high density integrated circuits these small devices are able to offer almost all the functionalities of desktop computers. Small size and low-cost of SBCs have made them a strong competitor in the computing market.

The popularity of SBCs has increased with the emergence of companies that produce commercially affordable devices such as Raspberry Pi [1], BeagleBone [2], and Arduino [3]. A series of functionalities such as the embedded wireless card, USB, and Ethernet port further justified the popularity of SBCs. Moreover, mass production reduced the price of each unit and made it more affordable for general purposes. The pocket-sized Raspberry Pi (only as one example) has become the third most popular computer in less than five years [4]. Distribution of over 10 million devices shows the currency of this device [5] in recent years.

Literatures have explored the use of SBCs in different industries (e.g., IoT [6, 7], health care [8, 9], and cloud [10, 11]). As an example, size and price have made SBC a major candidate to be used as the underlying hardware for IoT devices. Recently, IBM has specifically introduced integration of Watson

IoT platform with Raspberry Pi [12]. Sensly [13], a smart and portable air pollution sensor, Fridge [14], and Espresso[15], as two IoT-enabled appliances are a few examples of IoT devices using SBC.

While SBC had been the focus of researchers in several domains, less number of works have considered it in the realm of offensive security. In [16], authors have discussed possible usage of this device to launch different types of penetration tests to assess the security of a network. [17] uses Raspberry Pi as a Honeypot to detect SQL injection attacks and [18] deployed a Honeypot to simulate vulnerabilities and attract attackers. Hu et al. used this device as a distributed vulnerability assessment tool [19]. However, to the best of our knowledge, no previous research has been conducted to represent the potential misuse of SBC to threaten the security of a network.

Industrial espionage has existed long before the emergence of the Internet, however, modern technologies have facilitated theft of information. Prior to the emergence of commercially accessible SBC devices, access to spy gadgets and gears were limited. The finite number of producers and distributors available in the market made the traceability of these devices easy. For instance, there were only a limited number of companies to produce the physical key-loggers and market their devices publicly. Industrial espionage is now a major threat to the corporate world and can make long-term harm to companies. It has been reported that cyber-crime and economic espionage costs more than $445 billion annually in the world economy (almost 1 percent of global income) [20].

In this paper, we show how an attacker can deploy a full-fledged inexpensive attacking tool that can be mounted on networks if he has short-term physical access to the organization. We perform a set of experiments to demonstrate the possible harm that misemploying of SBC may cause. To show the feasibility of the attack, we use Raspberry Pi as a spying device to attack two real-world networks. We also provide possible directions for detection and prevention of such attacks.

SBC facilitates our attacks in several aspects. First, unlike the traditional form of insider attacks, the attacker does not need to plug a large computational device to the network. Large devices are more probable to be noticed while entering the organization or if left unattended for a long duration of time. In contrast, in our work, the attacker can benefit from the small size of SBC, and take them to premises and even leave them for a long period of time without grabbing the attention of security officers or employees. Second, in conventional attacks the attacker may exploit an insider machine to run a malicious software; however, such malwares may get detected by the machine's antivirus/malware tools or be noticed since they impact the performance of the local machine. SBC on the other hand, is an attacker owned device with no local anti-virus and malware tool installed and is not controlled by the system administrator. Third, using SBC, the attacker does not need to launch the attacks from outside the networks and, therefore, the possibility of detecting the attack by edge intrusion detection systems and firewalls reduces. All the mentioned points can play in favor of the attacker to smoothly perform the malicious activities without getting trapped.

**Contributions:** The detail contribution of our work is as follow.

1. *Attack Setup:* We turned a low-cost Raspberry Pi 3 (as an instance of SBC) into a spying gadget and plug it to a victim machine located in a real-world network to subliminally intercept the target machine's traffic. We installed Kali Linux on Raspberry Pi and loaded our gadget with multiple off-the-shelf malicious scripts and tools to exploit the victim. Our device is capable of launching different types of attacks including sniffing, spoofing, and man in the middle attacks.

2. *Experiment and Results*: We tested our gadget in two large size organization. Due to ethical considerations, we only intercepted the traffic targeted to the examiner's machine acting as the victim. We successfully launched several attacks including traffic sniffing and redirecting, and DNS service poisoning as a few examples of several possible attacks. We observed that both organizations were highly susceptible to the attacks launched from our small size spying gadget.

3. *Direction for Defense:* We conclude by reiterating defensive mechanisms that may be deployed to detect and prevent the suggested attacks. While threats are continually evolving and adapting to undermine protective measures, security measures as an ongoing set of practices and controls need to be updated to reduce the possible harms that new threats may cause.

**Paper Outline:** In Section 2, we present the potential attack scenario. Next, in Section 3, we present the experiment and its results. This is followed by Section 4, where we discuss possible defense mechanisms. Finally, we summarize our results and conclude our paper in Section 5.

## 2. ATTACK SETUP

In this section, we define the threat model and describe how an attacker can turn SBCs into an Espionage gadget.

### 2.1. Threat Model

In our study, the malicious adversary tries to launch network attacks against target victims in a large scale organization using an SBC. The attacker has short-term physical access to the organization, that is, he has access to enter the premises but is not allowed to carry advanced computational devices (e.g., laptop) into the building and leave it unattended.

In this scenario, the attacker enters the building carrying the pocket-size SBC (despite the organization's policy). After accessing the building, the attacker can plug the device in a hidden way to a computer or the network. The attacker may load the SBC with off-the-shelf tools and script to fully launch different type of attacks, including passive attacks (e.g., intercept and transfer private information to the outsiders or possibly store them locally for future offline access), or active attacks (e.g., DNS poisoning and man-in-the-middle attack) as will be discussed in Section 2.2.

Our hypothesis is that the attacker can plug the device in a subliminal way – owing to its size, and launch the attacks successfully, given poor network configuration and lack of optimal physical and network system countermeasures.

### 2.2. Turning Raspberry Pi into an Espionage Gadget

**Hardware Specification:** We selected Raspberry Pi [1] due to its popularity, affordability, and desired functionality it can provide. However, any other available SBC with network communication support can be chosen. We use the third generation of Raspberry Pi with the Quad Core 1.2GHz Broadcom BCM2837 64bit CPU, 1GB RAM, BCM43438 wireless LAN and Bluetooth Low Energy (BLE) on board, 4 USB 2 ports and Micro SD port for storing data.

**Physical Setup:** Several approaches may be taken to connect the Raspberry Pi device to the targeted network. If the attacker has unrestricted access to the victim's wireless network, he can connect and intercept the traffic. Similarly, the attacker may simply plug Raspberry Pi to any unattended network socket. However, commonly, administrators limit the number of available network sockets and restrict access to wireless networks and unplugged wired ones.

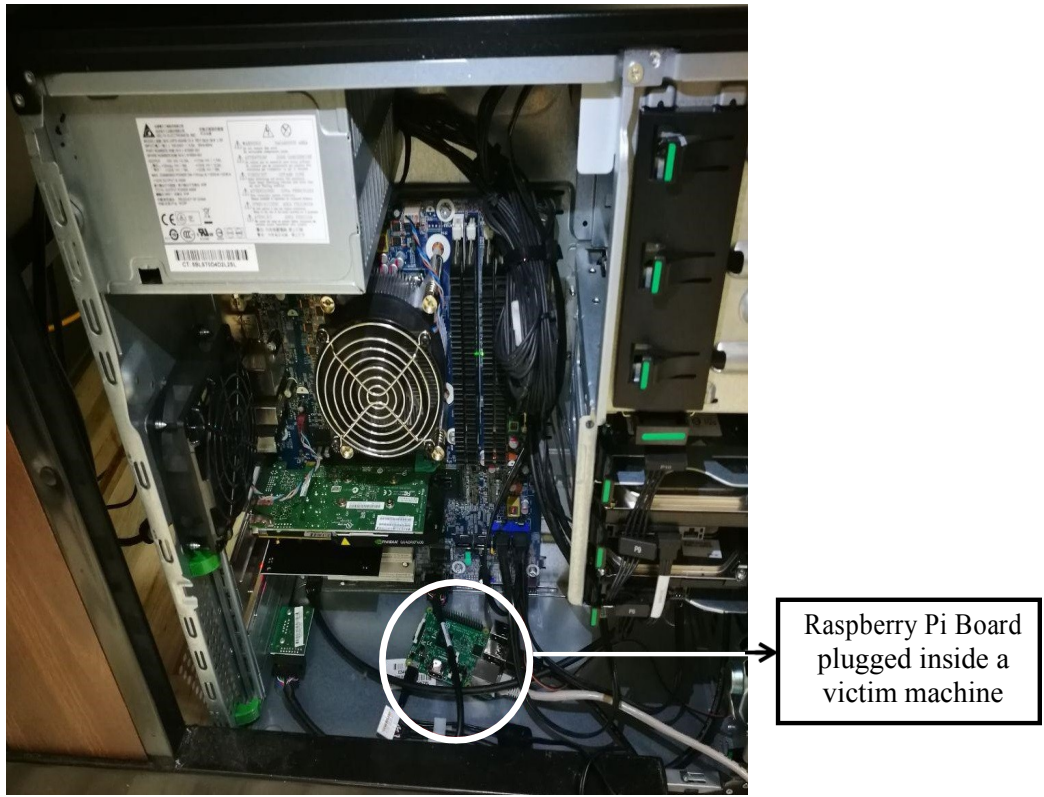

Fig. 1: Experiment Setup

Optionally, the attacker may connect the SBC to a victim machine to intercept and relay all the traffic targeted to the victim machine, while performing other network attacks. In this setting, the attacker puts the gadget between the target computer and the network (i.e., acting as a hub between the victim machine and the network). Hence, the device can use the network port primarily assigned to the target machine and intercept/relay all traffic designated to target machine. Figure 1 shows an example of such a setup used in our experiment.

Apart from the unavailability of network sockets, the main motivation behind this setup is to minimize the interception of traffic to one machine rather than the whole network due to ethical consideration (as will be fully discussed in chapter 3.1).

**Operating System:** We used Kali Linux [21] (A distribution of Linux for penetration testing) as an underlying platform for launching the attacks. Kali Linux offers two different pre-built versions that can be installed on Raspberry Pi: first, a light image, streamlined with the minimum tools; second, a full version that includes a Kali Linux full meta-package [22]. In this paper, we used the light version, which only encompasses some of the major Kali's application. The reason for installing the lighter version was that we found it more stable and also it helps the system to boot faster compared to the full meta-package.

**Network Attacking Tools:** After installing Kali Linux in Raspberry Pi, we installed a series of tools, which is needed to launch the attack. First, we installed the *Bridge-utills* package for sharing the Internet and making the Raspberry pi able to sniff the traffic somewhere between the targeted system and the network. This package lets Raspberry become a hub and accordingly helps to pass and intercept the traffic. Second, we installed the *tcpdump* packet analyzer to capture TCP/IP packets over a network. *tcpdump* provides the functionality of dumping the live packages and storing them into the dump files. Third, *Driftnet* [23], which listens to the network stream and picks images from TCP traffic. Finally, we installed *Ettercap* as a comprehensive suite for the man-in-the-middle attacks. This tool allows launching the DNS spoofing and redirecting the user to the attacker's website. Other tools can be loaded as per the attacker's requirement.

## 3. EXPERIMENT AND RESULT

### 3.1. Ethical Consideration

It has always been a dilemma whether computer network attacks are ethically correct. It is known that computer network attacks may harm an individual's and company's privacy, secrecy, reputation, and financial gains. However, if used in the correct way to aware the companies on possible vulnerabilities, it in fact, turns into a valuable tool helping to improve the security of the organizations. For more information about the ethics of computer network attacks please refer to [24].

Similar to any other network attack study, the purpose of our study and the affected target machine's defines the ethics of our work. In this research, we attempt to cast light on the fact that small SBCs have gained enough computational and communication power that they can be used as powerful attacking devices. We run several experiments to show how a real-world organization could be susceptible to attacks. In order to follow the ethical consideration, we designed our experiments so that we minimize the harm to the organizations by limiting the target of the attack to the experimenter's computer on the same network, as will be discussed in Section 3.2.

### 3.2. Experiment

To demonstrate the results of using the spying SBC in real world contexts, we conducted our experiments in two different organizations, an educational institution and a telecommunication company. We examined our scenario in a setting where the SBC device acts as a hub between a victim computer and the network. As mentioned in Section 3.1 the victim computer is the examiner's computer. Such a computer is a representative of other nodes in the same network with a similar setup and attack protective measures that are applied from a centralized network administration system.

Our SBC is able to capture all the traffic with the victim machine as the destination. Depending on the type of the attack, the attacker may store the traffic, transfer it to an external destination, or relay it to the victim (with or without manipulation). Note that the device is in fact capable of launching other types of active or passive attacks on the whole network, however, to eliminate the harm to the organization network, we pick attacks that only impact the victim machine. Figure 2 shows the attack overview where the attacker can mount the spying device to one or more nodes to launch passive and active targeted or holistic attacks.

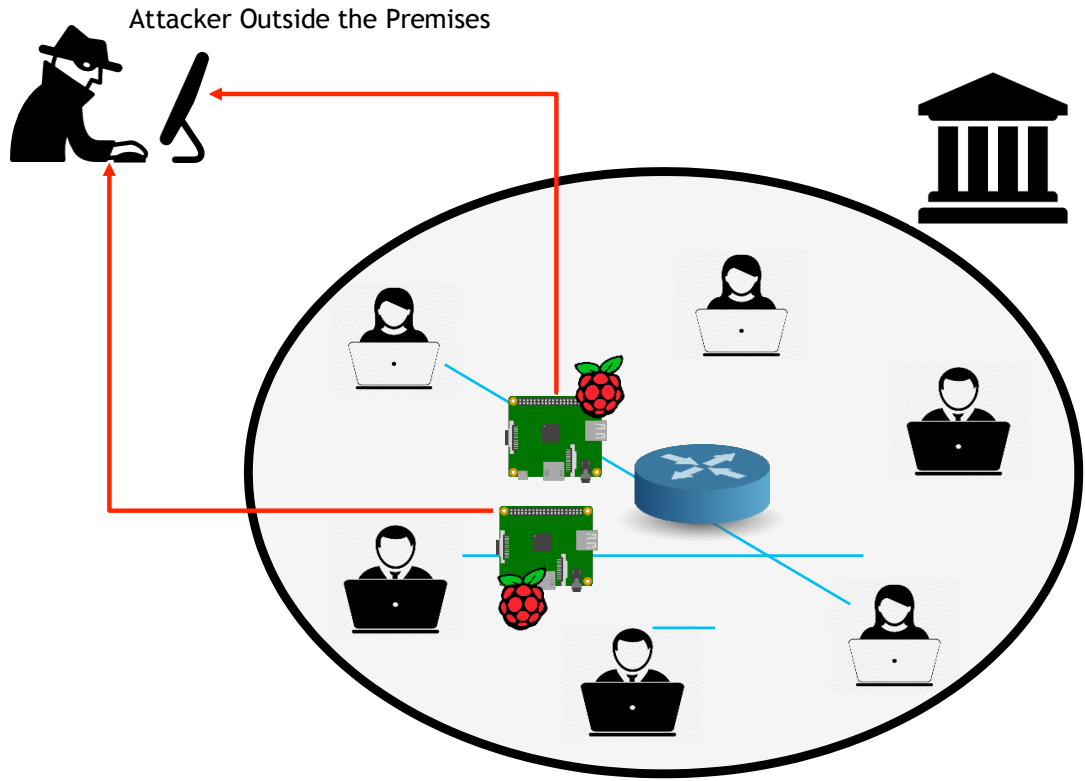

Fig. 2: Attack overview

**Transferring Data to the attacker:** As discussed in Section 2.1, the attacker has a limited, short-term access to the target company premises. Thereby, he needs to follow a mechanism to obtain his accumulated data. The attacker may store the data on the device and later on collect the device and analyze the stored data offline. Optionally, the attacker can use the device to access the data online by making the device to transfer the data in real-time or upon attacker's request. For the purpose of transferring the data to the attacker, we took two approaches: 1) emailing the collected data to the attacker and 2) storing the collected data on a cloud-based storage.

In the first option, we setup an email account on Gmail that is used by the attacker to receive the data stored on Raspberry Pi. We developed a Python program that creates an email with the Raspberry Pi data as a file attachment and sends it to the attacker email address every 60s (the timing is configurable). In the second option, we created a Dropbox account and a Dropbox API App under the same Dropbox account that can access files. We created a Dropbox directory to store files and gave the app access to this folder. We then developed a Python application that uses Dropbox library and uploads locally stored files (e.g., dump files) to Dropbox using an access token generated on Dropbox account.

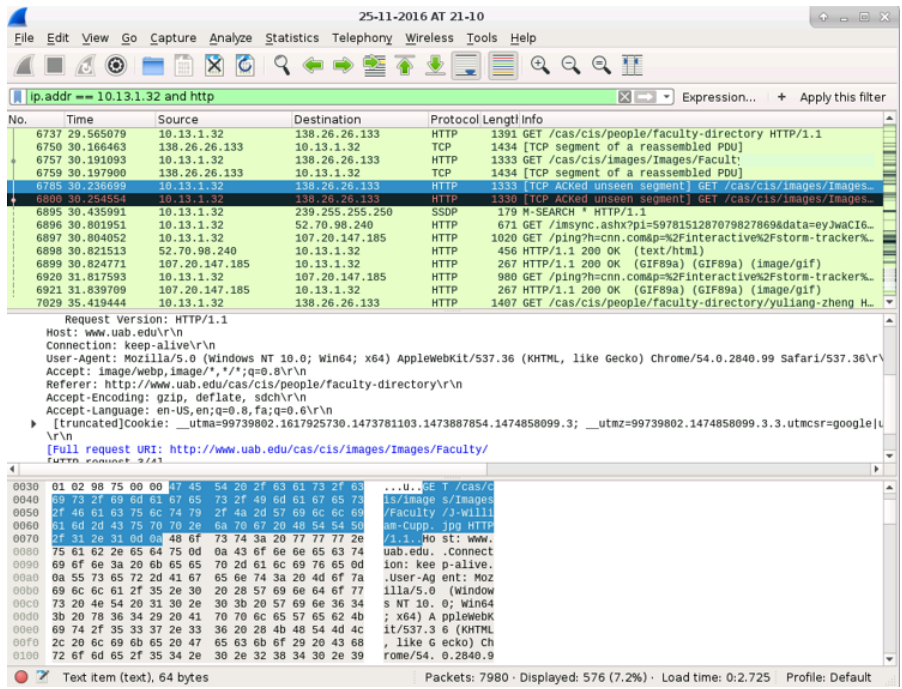

Fig. 3: Viewing intercepted traffic using Wireshark

**Intercepting Network Traffic:** For the purpose of sniffing network traffic, we used *tcpdump*, as a common command-line monitoring and packet analyzer tool. Using *tcpdump* network administrators are able to acquire network traffic for future debugging. *tcpdump* is a light-weight application and therefore is suitable to be loaded on Raspberry Pi.

We accessed the *tcpdump* data remotely and not from the Raspberry Pi device itself. Hence, we stored the dump data in a local file. Since the size of the file could become large, we developed a Python program that runs *tcpdump* and loads the data in one single text file in certain intervals. The stored files can be sent to the attacker using the transferring program we explained earlier in this section. The naming of the files are sequential so that the attacker can access them in the order they were generated.

Using *tcpdump* we were able to sniff and dump the data related to a user's visited website. The collected dump file can be analyzed using packet analyzers. Wireshark [25] is one of the most prominent network protocol analyzer tools. Figure 3 displays the *Wireshark* app debugging the obtained data from *tcpdump*. As can be seen in the figure the attacker can access the victim's visited website (or any other sensitive information send or received by the target machine).

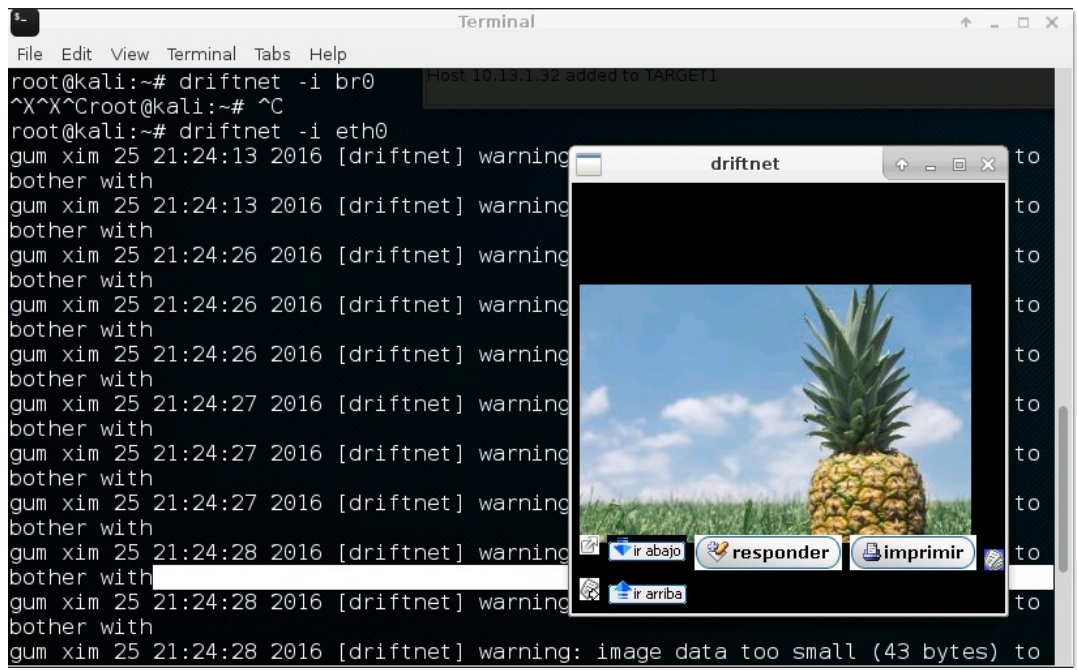

Fig. 4: Intercepted image viewed by victim using Driftnet

**Intercepting Visited Images:** *Driftnet* is an invasive application that listens to network traffic and picks MPEG audio streams and images from the TCP traffic. Using *Driftnet,* we were able to sniff user's viewed images. Figure 4 displays a visited image intercepted by *Driftnet*.

**Spoofing Attack:** Spoofing attacks express the situation when an attacker attempts to masquerades himself as another host or server by falsifying data to attain illegitimate information such as user's credentials. The attack occurs since TCP/IP does not provide any protection mechanism against spoofing to authenticate source and destination of a message. Thereby, the protocol is vulnerable to attack, while no extra verification measure is placed. The vulnerability, therefore, can be utilized to leverage man-in-the-middle attacks. There are several types of spoofing attacks, including, IP Address Spoofing, ARP Spoofing, and DNS Server Spoofing Attack.

To simulate these kinds of attacks, we used *Ettercap*. This tool allows the attacker to launch a variety of spoofing attacks to launch Man-in-the-Middle attack, and page or user redirection to a counterfeit host or service. During our attack we found both networks to be sustainable to different types of spoofing attacks. Figure 5 displays DNS spoofing attack as an example of attacked possible through Ettercap. In this experiment, we faked the DNS server for the targeted machine to redirect it to the attacker's website. We created a counterfeit website as shown in Figure 5 and by mapping the desired DNS to faked address, redirected the user to a counterfeit website (here an imitated Facebook page).

**Other Attacks:** In the case of illegitimate access to company premises, our gadget can potentially be used to launch a variety of attacks. Since the attacker using current implementation has turned to an insider node, the potential protective mechanisms such as firewall have already been bypassed and thereby the attack surfaces are limitless. For instance, while network foot-printing in a vast number of cases might not be viable from outside, it is more likely to work from the inside network using our gadget. For instance, we further expanded our test to observe the reflection of our target network against a Dos attack. To run this test, we developed a simple packet generator script which randomly sends UDP packets to the random UDP ports. By adding 5 instances of our gadget to our test environment, we noticed a significant amount of load on the network. Clearly, on a larger scale with more devices attached to a network, as a distributed decentralized attack, can result in a significant UDP storm and accordingly network downtime.

## 4. DISCUSSION AND POSSIBLE DEFENCE MECHANISEMS

As discussed, the small size of the computationally powerful SBC allows the attacker to leave it in the network (temporarily or permanently) without getting noticed. One possible source of attack seems to be lack of enough physical security measures that permit the attacker to enter the targeted building while carrying such device. However, organizations may be reluctant to enforce policies for entering digital devices to the buildings due to the prevalence of these devices. A more feasible approach is to deploy best practices in network security to detect and perhaps prevent attacks from unmanaged devices inside the network. In this section, we discuss possible attack origins and protective mechanisms that can help to safeguard companies' digital assets, and avert such data breaches.

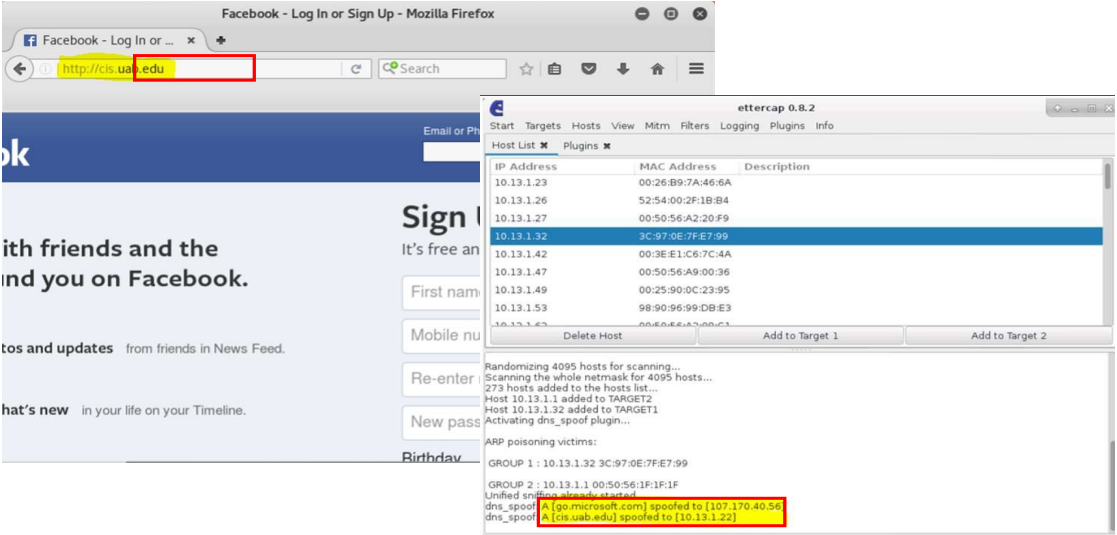

Fig. 5: DNS spoofing and redirecting user

### 4.1. Physical Security

Organizations try to improve their security maturity and the resistance to attacks using different types of protective measures. Limiting accesses and using physical security is one way of deterring malicious outsiders. Physical security as one of the crucial components of security is being taught in the majority of related educational materials. A variety of researches investigated the physical security from different angle and context. Several literatures discussed the influence and effectiveness of the physical security in the industry [26, 27]. The traditional definition of the physical security was limited to protecting assets against perils such as flood, fire, and burglary. The modern definition of the term describes methods and techniques that are utilized to prevent or deter unauthorized physical accesses and thereby safeguard information.

Physical security as a barrier placed around the organization digital assets [28] provides an additional layer of security against malicious outsider and adversary insider. Many of the big (and even medium size) organizations control their doorways to avoid entrance of digital devices (e.g., laptops, cameras and networking devices) to minimize the risk associated with launching attacks and theft of information. However, the development of mobile phones and small computational devices seems to have made physical security more challenging. In this experiment, we showed that the attacker bypassed the physical security measures in the two mentioned organizations and took the Raspberry Pi device inside the network. The result of our experiments delineates the importance of revising the traditional definition of physical security and reconsidering the efficiency of it against such pocket-sized threats.

### 4.2. Network Security

**Network Configuration.** First protective mechanism lies on system configuration. At the very first step our suggested attacks were successfully launched as there was no mechanism to prevent an untrusted device to join the network. More specifically the result of this attack was promising due to the misconfiguration in upstream switch/router which allowed SBC to appear as a switch for redirecting traffic in the network. The attack can be prevented or be far less harmful if *"unauthorized switches"* configuration [29] was active in the upper layers of network. Means that the target network could be impervious to attack by activating *"unauthorized switches"*.

**Attack Detection Script.** While proper configuration can ideally prevent this type of attacks, depending on the environment conditions such configuration might not be applicable. This specifically can happen for the cases where a company involves considerable commuting in devices which makes the device tracking tough or even impossible to follow. Therefore, in this section, we suggest use of scanning tools which can detect the presence of sniffing devices in the network.

We assume that the victim network is a switched Ethernet and traffic is not transferred on a shared media (e.g., hub, bus). In a switched network, there are several methods to sniff the network traffic. The simplest sniffer is set by configuring the network card into promiscuous mode and sniffing all traffic matching a targeted MAC address. Another type of sniffing relies on ARP poisoning, in which the attacker poisons the ARP cache and links the IP address of a legitimate user to its own MAC address, therefore, any packet intended to the IP address will reach the victim.

Several tools have already been developed to detect if a node on a network intercepts the traffic. *"Nmap"* [30] is a network security tool created to scan the network for administration purposes. *"Nmap"* sends a packet to the target host and receives the responses that can be used to assess different security parameters. *"Nmap"* has developed a script based on the method suggested in [31] to detect whether a network card is in promiscuous mode. This method creates fake ARP request packets that are sent to every node on the network. Nodes that are set in promiscuous mode respond to this ARP requests while other genuine nodes block the request. Another tool created by researchers is AntiSniff [32] that uses a variety of attacking techniques to not only recognize sniffing devices with the Ethernet cards in promiscuous mode but also to detect active attackers. Using such tools would heavily help the networks to detect and prevent attacks launch from an insider node.

## 5. CONCLUSION

Prior to the emergence of SBCs, to produce a spying gadget, attackers needed to have an advanced technical knowledge to build the hardware and program it according to the requirements, or to plug laptops into target networks to launch attacks, which could have been spotted. However, SBC devices made it easier for attackers to run networking attacks without getting noticed.

In this study, we have witnessed how a malicious adversary is able to launch an attack against an organization using off the shelf devices and tools. We showed how one can turn a small size SBC into a full-fledged network attacking gadget. We configured and installed the device in two organizations and showed examples of the possible network attacks. Given the popularity of SBC devices and their powerful resources we suggested deploying network monitoring schemes to detect and prevent such malicious network activities.